\documentclass[twocolumn,pra,aps,superscriptaddress,floatfix]{revtex4}
\usepackage{amssymb}
\usepackage{amsfonts}
\usepackage{amsmath}
\usepackage{graphicx}

\begin{document}

\title{Detecting quantumness with generalized Loschmidt echoes}
\date{\today }
\author{Cecilia Cormick}
\affiliation{Instituto de F\'isica de la Facultad de Ingenier\'ia, Universidad de la
Rep\'ublica, Julio Herrera y Reissig 565, Montevideo, Uruguay}
\author{Adri\'an A. Budini}
\affiliation{Consejo Nacional de Investigaciones Cient\'ificas y T\'ecnicas (CONICET),
Centro At\'omico Bariloche, Avenida E. Bustillo Km 9.5, (8400) Bariloche,
Argentina, and Universidad Tecnol\'ogica Nacional (UTN-FRC), Fanny Newbery
111, (8400) Bariloche, Argentina}

\begin{abstract}
Loschmidt echoes are a well-established method to characterize dynamical
properties such as quantum chaos and sensitivity to perturbations. Here we
present a straightforward generalization that identifies non-classical
dynamical behavior. The generalized echoes involve four forward-backward
time propagators. If these propagators do not commute, one can readily
identify signatures of this property in the relations between echoes. As an
example, detection of intrinsic non-commuting quantum features is analyzed
in critical dynamics of a spin chain. Interestingly, this phenomenon is also
observed for Gaussian dynamics and high temperature limits which are
generally regarded as classical.
\end{abstract}

\maketitle

\section{Introduction}

The Loschmidt echo (LE) is a tool to measure to what extent a quantum system
returns to its initial state after a certain evolution \cite{Zhang_1992,
Pastawski_1995, Levstein_1998, Jalabert_2001, Jacquod_2001, Cucchietti_2003}
. This process is typically composed of a standard \textquotedblleft forward
evolution\textquotedblright\ followed by an approximate \textquotedblleft
backward evolution\textquotedblright\ corresponding to a perturbation of the
inverse of the first step. The LE is used, both theoretically and
experimentally, to study chaotic behavior, decoherence, information
scrambling, criticality and sensitivity to perturbations in quantum systems 
\cite{Wisniacki_2003, Prosen_2006, Quan_2006, Cormick_2008, Calvo_2008,
Yan_2020, Sanchez_2020, Hasegawa_2021,scholar}.

In this work we present a simple extension of the Loschmidt echo, producing
quantities that involve one more step of forward/backward evolution. These
Generalized Loschmidt echoes (GLEs) allow a straightforward detection of
intrinsic quantum features. Our protocol shares some similarities with other
approaches for the characterization of dynamics, such as two-dimensional
spectroscopy \cite{Ernst_1989, Mukamel_1995} or alternative sensing
strategies \cite{Degen_2017, Kuffer_2025}, that also explore the system's
response to different kinds of perturbations, pulses or control
Hamiltonians. The appealing feature of the GLEs is that they inherit the
conceptual and practical simplicity of the standard LE, while being able to
signal the presence of quantum-mechanical features in the dynamics and also
to characterize statistical features of their fluctuations in time.


It is important to note that the distinction of what is \textquotedblleft
genuinely quantum\textquotedblright\ is a question almost as old as quantum
mechanics itself. Its answer has varied significantly along time and depends
largely on the context. For instance, quantumness can be defined in
inequivalent ways in terms of negativity of the Wigner function \cite%
{Wigner,Combescure_2012, Hudson_1974, Soto_1983} or of violation of
Bell-like inequalities \cite{Bell,Nha_2004, Lezama_2023}. Classical
non-invasive measurability leads to Leggett-Garg inequalities \cite{Leggett,
Nori} which in turn were analyzed recently in the context of quantum
non-Markovianity \cite{Milz,Horo,Budini}. Independently of many other
consistent definitions, within the present study we understand
\textquotedblleft classical dynamics\textquotedblright\ as those in which
all propagators commute. Thus, a signal of non-vanishing commutators is
interpreted as a quantum feature. Quantumness is thus associated with
incompatibility of the generators of the evolution \cite{Gahne_2023}.

In the following we will introduce our generalizations of the Loschmidt echo
and explain how different assumptions about the dynamics, such as
classicality or form of the time correlations, lead to certain relations
between GLEs. We will also discuss the application to two relevant examples
given by critical spin chains and thermal harmonic environments.

\section{Generalized Loschmidt echoes}

The LE corresponds to the overlap between two states of a possibly complex
quantum system. Each state is chosen to be given by the same initial state,
which here we label $|\mathcal{B}\rangle $, time-evolved with a different
Hamiltonian. Denoting the corresponding propagators as $\mathbb{G}%
_{t_{f},t_{i}}^{(s)}$ with $s=\pm $ and where $t_{i}$\ and $t_{f}$ are the
initial and final times respectively, we define the LE as\ \cite{EcoModulo}%
\begin{equation}
\mathrm{E}(\tau )\equiv \langle \mathcal{B}|\mathbb{G}_{\tau ,0}^{(-)\dag }%
\mathbb{G}_{\tau ,0}^{(+)}|\mathcal{B}\rangle .  \label{StandardLE}
\end{equation}%
The product $\mathbb{G}_{\tau ,0}^{(-)\dag }\mathbb{G}_{\tau ,0}^{(+)}$\ is
usually named the echo propagator . The definition above can be generalized
to the case of initial mixed states in a straightforward manner.

The expression for $\mathrm{E}(\tau )$ only involves one time interval with
two different propagators. The GLEs proposed in this contribution read 
\begin{equation}
\mathrm{E}_{\tilde{s}s}(t,\tau )\equiv \langle \mathcal{B}|\mathbb{G}%
_{t,0}^{(\tilde{s})\dag }\mathbb{G}_{t+\tau ,t}^{(-)\dag }\mathbb{G}_{t+\tau
,t}^{(+)}\mathbb{G}_{t,0}^{(s)}|\mathcal{B}\rangle ,  \label{GeneralizedLE}
\end{equation}%
where $s,\tilde{s}\in \{+,-\}$. This definition involves four propagators
over two different consecutive time intervals, with durations $t,\tau \geq
0. $ The intermediate product $\mathbb{G}_{t+\tau ,t}^{(-)\dag }\mathbb{G}%
_{t+\tau ,t}^{(+)}$ is the standard echo operator shifted in time, which in
turn is contracted with the (time-evolved) states given by $\mathbb{G}%
_{t,0}^{(s)}|\mathcal{B}\rangle $ and $\langle \mathcal{B}|\mathbb{G}%
_{t,0}^{(\tilde{s})\dag }.$ Consequently, the GLEs could be implemented with
standard experimental techniques associated to the standard case.

By construction, the four GLEs satisfy $\mathrm{E}_{\tilde{s}s}(0,\tau )=%
\mathrm{E}(\tau ).$ In addition, $\mathrm{E}_{++}(t,0)=\mathrm{E}%
_{--}(t,0)=1,$ while $\mathrm{E}_{+-}(t,0)=\mathrm{E}^{\ast }(t)$ and $%
\mathrm{E}_{-+}(t,0)=\mathrm{E}(t).$ Using the divisibility of the unitary
propagators, the four GLEs can be rewritten as 
\begin{subequations}
\label{ExplicitGLE}
\begin{eqnarray}
\mathrm{E}_{++}(t,\tau ) &=&E(\tau |t)+\langle \mathcal{B}|[\mathbb{G}%
_{t,0}^{(+)\dagger },\mathbb{G}_{t+\tau ,t}^{(-)\dagger }]\mathbb{G}_{t+\tau
,0}^{(+)}|\mathcal{B}\rangle ,\ \ \ \ \  \\
\mathrm{E}_{--}(t,\tau ) &=&E(\tau |t)-\langle \mathcal{B}|\mathbb{G}%
_{t+\tau ,0}^{(-)\dag }[\mathbb{G}_{t,0}^{(-)},\mathbb{G}_{t+\tau ,t}^{(+)}]|%
\mathcal{B}\rangle , \\
\mathrm{E}_{\pm \mp }(t,\tau ) &=&\langle \mathcal{B}|\mathbb{G}_{t,0}^{(\pm
)\dag }\mathbb{G}_{t+\tau ,t}^{(-)\dag }\mathbb{G}_{t+\tau ,t}^{(+)}\mathbb{G%
}_{t,0}^{(\mp )}|\mathcal{B}\rangle ,
\end{eqnarray}%
where for shortening the notation we introduced $\mathrm{E}(\tau |t)\equiv
\langle \mathcal{B}|\mathbb{G}_{t+\tau ,t}^{(-)\dag }\mathbb{G}_{t+\tau
,t}^{(+)}|\mathcal{B}\rangle $, which is the standard LE but with a time
displacement. Of central relevance for the present proposal is the fact that
the difference between the first two GLEs is explicitly sensitive to
non-vanishing commutators of propagators, that is, a purely quantum
property. The two remaining echoes will be shown to be associated with the
time correlations of the dynamics. Notice that the last of the echoes
corresponds simply to the standard LE over the total time interval, that is, 
$\mathrm{E}_{-+}(t,\tau )=\mathrm{E}(t+\tau )$.

\subsection{Loschmidt echoes and classical noise}

Let us now assume that the dynamics are classical, in the sense that all
commutators between propagators are negligible. In this case, there exists a
basis of eigenstates which is common to both (in general time-dependent)
Hamiltonians. Then, the LE [Eq.~(\ref{StandardLE})] can be written as 
\end{subequations}
\begin{equation}
\mathrm{E}(\tau )=\langle \exp [-i\textstyle\int_{0}^{\tau }\xi (t^{\prime
})dt^{\prime }]\rangle ,
\end{equation}%
where $\xi (t)\equiv \xi ^{(+)}(t)-\xi ^{(-)}(t)$ with $\xi ^{\pm }(t)$ the
corresponding energies, while $\langle \cdots \rangle $ denotes the
expectation value. This expression can be interpreted equally as an average
over different stochastic realizations, with a probability distribution
determined by the initial state and the basis diagonalizing the
Hamiltonians. Conversely, any echo that can be interpreted in this way can
be associated with a classical noise $\xi (t).$

Similarly, from Eq.~(\ref{ExplicitGLE}) in the scenario of classical noise
the GLEs become 
\begin{subequations}
\begin{eqnarray}
\mathrm{E}_{++}(t,\tau ) &=&\langle \exp [-i\textstyle\int_{t}^{t+\tau }\xi
(t^{\prime })dt^{\prime }]\rangle , \\
\mathrm{E}_{--}(t,\tau ) &=&\langle \exp [-i\textstyle\int_{t}^{t+\tau }\xi
(t^{\prime })dt^{\prime }]\rangle , \\
\mathrm{E}_{\pm \mp }(t,\tau ) &=&\langle \exp [-i\textstyle\int_{t}^{t+\tau
}\xi (t^{\prime })dt^{\prime }\pm i\textstyle\int_{0}^{t}\xi (t^{\prime
})dt^{\prime }]\rangle ,\ \ \ 
\end{eqnarray}%
The first two lines are identical so that 
\end{subequations}
\begin{equation}
\mathrm{E}_{++}(t,\tau )=\mathrm{E}_{--}(t,\tau )\equiv \mathrm{E}(\tau
|t)\quad \mathrm{(classical~noise).}
\end{equation}%
Furthermore, for stationary noise, that is, when the noise statistics are
invariant under time translations, the first two GLEs are equal of the
standard echo, $\mathrm{E}(\tau |t)\to \mathrm{E}(\tau ).$

While the properties above are common to all classical noise models, the
structure of the echoes depends on the noise statistics. Particularly simple
expressions can be obtained in two limit cases for the time correlations. In
the \textit{white noise case,\ }$\langle \xi (t)\xi (t^{\prime })\rangle
\propto \delta (t-t^{\prime })$, leading to 
\begin{equation}
\mathrm{E}_{+-}(t,\tau )=\mathrm{E}(\tau )\mathrm{E}^{\ast }(t),\ \ \ \ \ 
\mathrm{E}_{-+}(t,\tau )=\mathrm{E}(\tau )\mathrm{E}(t).  \label{Ecowhite}
\end{equation}%
In this limit $\mathrm{E}(\tau )$ and quantities $\mathrm{E}_{\pm \pm
}(t,\tau )$ also fulfill a semigroup property (exponential decay). On the
other hand, in an \textit{infinite time correlation limit}, or equivalently
a random frequency model, for each realization one has $\xi (t)=\xi $ with $%
\xi $ a time-independent random variable, so that%
\begin{equation}
\mathrm{E}_{+-}(t,\tau )=\mathrm{E}(\tau -t),\ \ \ \ \ \ \ \ \mathrm{E}%
_{-+}(t,\tau )=\mathrm{E}(\tau +t).  \label{EcoRandomFrequency}
\end{equation}%
Further examples can be analyzed once a noise statistics is assumed; a
particularly relevant case is that of \textit{Gaussian noise}~\cite{kampen}
(see Appendix~\ref{GaussNoise}).

\subsection{Dynamical properties encoded in the GLEs}

In the present proposal, signatures of intrinsic quantum dynamics are
associated with non-commuting time propagators. From our previous discussion
we conclude that the difference $|\mathrm{E}_{++}(t,\tau )-\mathrm{E}%
_{--}(t,\tau )|$ measures departures with respect to a classical noise
model. Thus, this difference between GLEs quantifies intrinsic quantum
features of the dynamics. Its magnitude depends on each specific dynamics.
In addition, $|\mathrm{E}_{+-}(t,\tau )-\mathrm{E}(\tau )\mathrm{E}^{\ast
}(t)|$ can be used to measure departures from the white-noise limit, and $|%
\mathrm{E}_{+-}(t,\tau )-\mathrm{E}(\tau -t)|$ from a random frequency model
with infinite time correlation. In the following these main points are
explicitly illustrated through two different paradigmatic examples.

\section{Quantum departures from a random-frequency model}

For simplicity we now assume that the Hamiltonians $H^{(\pm )}$ that
generate the propagators are time-independent, and that the initial state $|%
\mathcal{B}\rangle $ corresponds to an eigenstate of $H^{(-)}$. Then, the LE
can be read as a statistical superposition of oscillatory terms with weights
given by the change of basis between the two Hamiltonians. Thus, one can
interpret the standard LE as resulting from a simple random-frequency
classical model. In contrast, under these assumptions one has $\mathrm{E}%
_{++}(t,\tau )=\mathrm{E}(\tau )+\langle \mathcal{B}|[\mathbb{G}%
_{t}^{(+)\dag },\mathbb{G}_{\tau }^{(-)\dag }]\mathbb{G}_{t+\tau }^{(+)}|%
\mathcal{B}\rangle ,$ while $\mathrm{E}_{--}(t,\tau )=\mathrm{E}(\tau ).$
Furthermore, $\mathrm{E}_{+-}(t,\tau )=\mathrm{E}(\tau -t)+\langle \mathcal{B%
}|[\mathbb{G}_{t}^{(+)\dag },\mathbb{G}_{\tau }^{(-)\dag }]\mathbb{G}_{\tau
}^{(+)}\mathbb{G}_{t}^{(-)}|\mathcal{B}\rangle ,$ and $\mathrm{E}%
_{-+}(t,\tau )=\mathrm{E}(\tau +t).$ Thus, the differences with respect to a
random frequency model [Eq.~(\ref{EcoRandomFrequency})] are proportional to
commutators of the underlying propagators, that is, this departure is
intrinsically quantum.

As an example we study the GLEs corresponding to an Ising chain with the two
time propagators corresponding to scenarios with different transverse fields 
$\lambda ^{(\pm )}$, as in \cite{Quan_2006}. The propagators are $\mathbb{G}%
_{t,t_{0}}^{(\pm )}=\exp [-iH^{(\pm )}(t-t_{0})],$ where%
\begin{equation}
H^{(\pm )}=-J\,\sum_{j=1}^{N}\lambda ^{(\pm )}\sigma _{j}^{(x)}+\sigma
_{j}^{(z)}\sigma _{j+1}^{(z)}.  \label{H_PM_Chain}
\end{equation}%
Here $N$ is the number of spins, while $\sigma _{j}^{(x)}$ and $\sigma
_{j}^{(z)}$ are the Pauli operators associated with the $x$- and $z$%
-directions respectively and acting on the $j$-th spin. Taking periodic
boundary conditions, one identifies $j=N+1$ with $j=1$. For definiteness, we
assume that $|\mathcal{B}\rangle $ is the ground state of $H^{(-)}$. Using
the Jordan-Wigner transformation and a Fourier decomposition, the LE can be
obtained in an exact analytical way~\cite{Dziarmaga_2005}.

Because of translational invariance along the ring, the LE can be factorized
in the form $\mathrm{E}(\tau )=\prod_{k}\mathrm{E}_{k}(\tau ),$ where $%
\mathrm{E}_{k}(\tau )$ can be read as the LE associated with the fermionic
modes of quasimomentum $\pm k$ (see Appendix \ref{Ising}). In the end, one
obtains an echo with the following structure: 
\begin{equation}
\mathrm{E}(\tau )=\Big\langle\exp \Big(-i\tau \sum_{k}\xi _{k}\Big)%
\Big\rangle=\prod_{k}\langle \exp (-i\tau \xi _{k})\rangle .  \label{LESpin}
\end{equation}%
Thus, the LE can be represented as a classical average over the random
frequency $\xi =\sum_{k}\xi _{k},$ where the probability density $P_{k}(\xi
_{k})$ of each statistically independent random variable $\xi _{k}$ is%
\begin{multline}
P_{k}(\xi _{k})=\delta \Big(\xi _{k}-\mathrm{E}_{k}^{(+)}\Big)\sin
^{2}(\Delta _{k}/2) \\
+\delta \Big(\xi _{k}-\mathrm{E}_{k}^{(-)}\Big)\cos ^{2}(\Delta _{k}/2).
\label{Probs}
\end{multline}%
In this expression, $\delta (x)$ is the Dirac delta function, $\mathrm{E}%
_{k}^{(\pm )}\equiv \epsilon _{k}^{(-)}\pm \epsilon _{k}^{(+)}$ is defined
in terms of the quasiparticle energies $\epsilon _{k}^{(\pm )}$ of each
quasimomentum block, while $\Delta _{k}=\phi _{k}^{(+)}-\phi _{k}^{(-)}$ is
determined by the angles $\phi _{k}^{(\pm )}$ of the Bogoliubov
transformations diagonalizing the Hamiltonian~(Appendix \ref{Ising}). Hence,
even though the underlying dynamics can cross a quantum phase transition~%
\cite{Sachdev_2011}, the LE, independently of the regime, can be identified
with a classical random-frequency noise. On the other hand, the GLEs can
distinctly indicate departures with respect to this kind of decoherence
model.

Following a similar procedure as for the standard LE, the GLEs can also be
obtained in an exact analytical way, and they also factorize as:%
\begin{equation}
\mathrm{E}_{\tilde{s}s}(t,\tau )=\prod_{k}\mathrm{E}_{k}^{\tilde{s}s}(t,\tau
).
\end{equation}%
Here, $\mathrm{E}_{k}^{\tilde{s}s}(t,\tau )$ is the GLE associated with each
independent pseudomomentum space (Appendix \ref{Ising}). Because of the
non-vanishing commutators between propagators, these objects cannot be cast
in terms of the classical random-frequency model associated with the LE
[Eqs.~(\ref{LESpin})-(\ref{Probs})].

As an illustration, in Fig.~1 we plot the differences $|$\textrm{$E$}$%
_{++}(t,\tau )-\mathrm{E}_{--}(t,\tau )|$ and $|\mathrm{E}_{+-}(t,\tau )-%
\mathrm{E}(\tau -t)|$, which signal deviations from purely classical noise
and from classical noise with infinite time correlation, respectively. The
inset shows the LE decay. For the parameters chosen, which lie very close to
the phase transition, this decay can be fit by an exponential function, $|%
\mathrm{E}(\tau )|\approx \exp [-a\tau ].$ This simple LE can be described
as resulting from a random frequency model with a Lorentzian distribution.
In contrast, the GLEs detect the quantum character of the dynamics. 
\begin{figure}[t]
\includegraphics[bb=47 872 726
1140,angle=0,width=8.5cm]{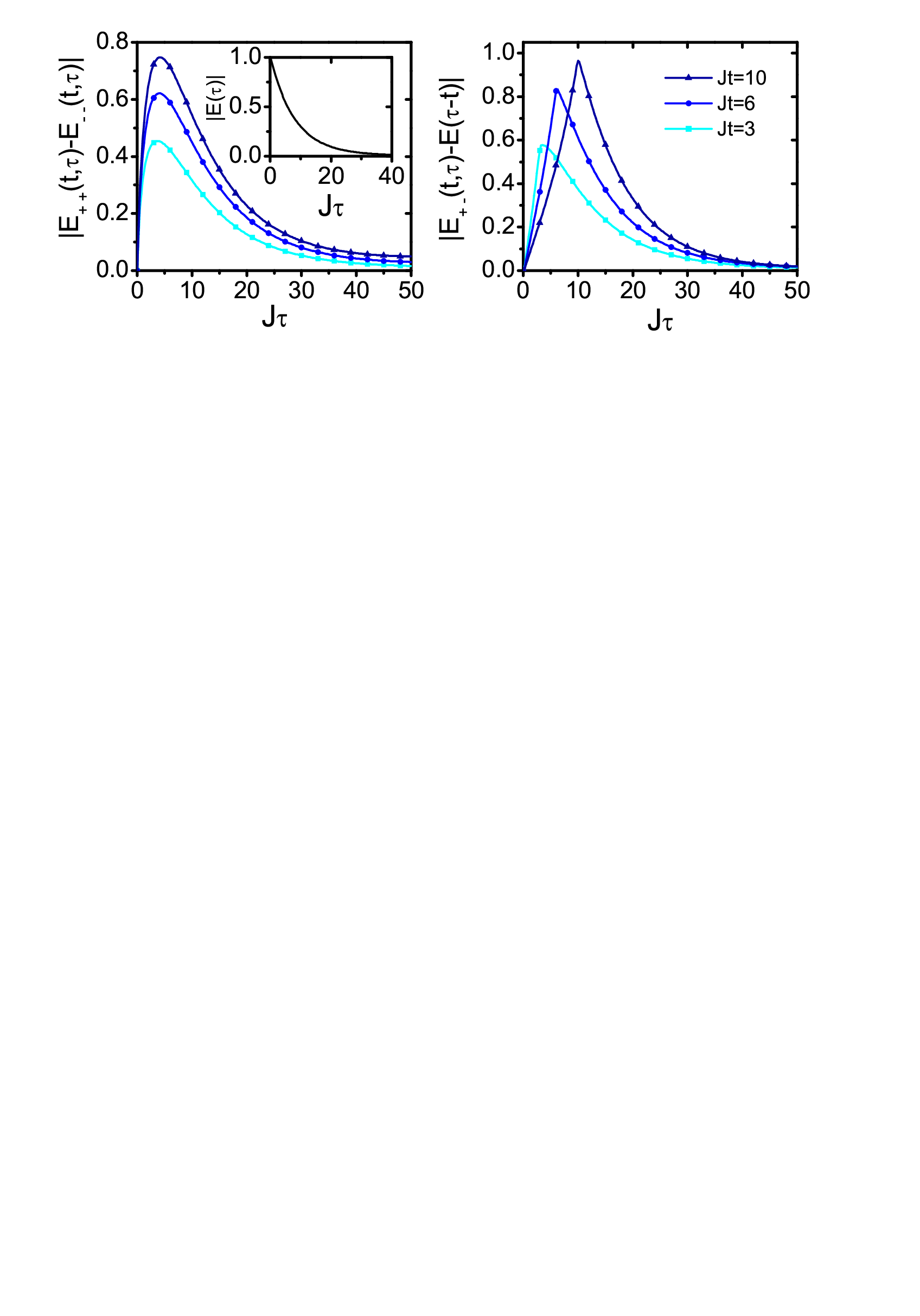}
\caption{Non-classicality quantifier $|\mathrm{E}_{++}(t,\protect\tau )-%
\mathrm{E}_{--}(t,\protect\tau )|$ (left panel) and departure from a random
frequency model $|\mathrm{E}_{+-}(t,\protect\tau )-\mathrm{E}(\protect\tau %
-t)|$ (right panel) for the Ising chain dynamics of Eq.~(\protect\ref%
{H_PM_Chain}). The parameters are $N=5000,$ $\protect\lambda ^{(+)}=1.011$
and $\protect\lambda ^{(-)}=1.001.$ The inset shows the standard LE, which
can be fit by an exponential decay.}
\end{figure}

\section{Quantum vs. classical Gaussian noises}

Here we consider another archetypical model of a quantum system, namely a
set of non-interacting quantum harmonic oscillators with Hamiltonian $%
H=\sum_{k}\omega _{k}b_{k}^{\dag }b_{k}.$ The perturbations are taken as
generalized linear force operators, which in interaction picture are of the
form: 
\begin{equation}
H^{\pm }(t)=\pm \sum_{k}(g_{k}e^{+i\omega _{k}t}b_{k}^{\dag }+g_{k}^{\ast
}e^{-i\omega _{k}t}b_{k})\,.  \label{bosonicHI}
\end{equation}%
Here, $|g_{k}|$ measures the strength of the force applied to the harmonic
oscillator with frequency $\omega _{k}.$ The resulting propagators $\mathbb{G%
}_{t,t_{0}}^{(\pm )}=\mathcal{T}\exp [-i\int_{t_{0}}^{t}dt^{\prime }H^{(\pm
)}(t^{\prime })]$, with $\mathcal{T}$ the time-ordering superoperator, are
displacement operators~\cite{Breuer_2002}, so that the dynamics are
analytically solvable with standard tools (see Appendix \ref{AppBoson}).
Furthermore, the calculation is simplified by the fact that the propagators $%
\mathbb{G}_{t,t_{0}}^{(\pm )}$ correspond to equal displacements in opposite
phase-space directions, and by the factorization of the echo in the
contributions corresponding to the different independent oscillators.

In this example, we will assume an initial condition corresponding to a
thermal state of the unperturbed Hamiltonian with inverse temperature $\beta 
$. The LE then reads%
\begin{equation}
\mathrm{E}(\tau )=\exp (-\gamma _{\tau }),  \label{EchoBosonic}
\end{equation}%
where 
\begin{equation}
\gamma _{\tau }=4\sum_{k}\frac{|g_{k}|^{2}}{\omega _{k}^{2}}[1-\cos (\omega
_{k}\tau )]\coth (\beta \hbar \omega _{k}/2).
\end{equation}%
The dimensionless decay function $\gamma _{\tau }$ also determines the GLEs
for $\tilde{s}=-s$. After some calculations they can be cast as:%
\begin{equation}
\mathrm{E}_{\pm \mp }(t,\tau )=\exp [-(\gamma _{\tau }+\gamma _{t}\pm \Gamma
_{t,\tau })],  \label{ECross}
\end{equation}%
where%
\begin{equation}
\Gamma _{t,\tau }=\gamma _{t}+\gamma _{\tau }-\gamma _{t+\tau }.
\end{equation}

The structure of the LE~(\ref{EchoBosonic}), as well as the one of the GLEs
in Eqs.~(\ref{ECross}) also emerges when the force is represented by
classical Gaussian noise with null mean, which leads to $\gamma _{\tau
}=\int_{0}^{\tau }dt_{2}\int_{0}^{\tau }dt_{1}f(|t_{2}-t_{1}|),$ where $%
f(|\tau |)$ is the stationary noise correlation (Appendix \ref{AppBoson}).
Thus, taking $f(|\tau |)=2\sum_{k}|g_{k}|^{2}\cos (\omega _{k}\tau )\coth
(\beta \hbar \omega _{k}/2)$ the echoes $\mathrm{E}(\tau )$ and $\mathrm{E}%
_{\pm \mp }(t,\tau )$ are indistinguishable from those resulting from
classical Gaussian noise. Departure with respect to a white noise limit is
set by $|\mathrm{E}_{+-}(t,\tau )-\mathrm{E}(\tau )\mathrm{E}^{\ast
}(t)|=\exp [-(\gamma _{\tau }+\gamma _{t})]|\exp (-\Gamma _{t,\tau })-1|.$

Nevertheless, this classical Gaussian noise representation is inconsistent
with the remaining two GLEs, with $\tilde{s}=s$:%
\begin{equation}
\mathrm{E}_{\pm \pm }(t,\tau )=\exp (-\gamma _{\tau }\pm i\Phi _{t,\tau }).
\label{Average4Guno}
\end{equation}%
The phase contribution can be written as%
\begin{equation}
\Phi _{t,\tau }=\phi _{t}+\phi _{\tau }-\phi _{t+\tau },
\end{equation}%
where%
\begin{equation}
\phi _{t}=4\sum_{k}\frac{|g_{k}|^{2}}{\omega _{k}^{2}}\sin (\omega _{k}t).
\end{equation}%
This phase is a quantum effect that has its origin in the lack of
commutativity between the propagators associated to $H^{\pm }(t).$ The
validity of a classical noise representation would imply $\Phi _{t,\tau }=0.$
The deviation from the classical noise equality, $|\mathrm{E}_{++}(t,\tau )-%
\mathrm{E}_{--}(t,\tau )|=2\exp (-\gamma _{\tau })|\sin (\Phi _{t,\tau })|,$
depends on the times considered, the spectral density, and the temperature.

For definiteness, we now evaluate a specific, ubiquitous example. We take a
continuous limit, $\sum\nolimits_{k}4|g_{k}|^{2}\to \int_{0}^{\infty
}d\omega J(\omega ),$ and choose an Ohmic spectral density with a Lorentzian
cutoff $J(\omega )=A\omega \gamma ^{2}/(\omega ^{2}+\gamma ^{2})\pi $. In
the high-temperature limit ($\beta \hbar \gamma \ll 1$) the rate can be
approximated as (Appendix \ref{AppBoson})%
\begin{equation}
\gamma _{t}\simeq \frac{A}{\beta \hbar \gamma }[\gamma t-(1-e^{-\gamma t})],
\end{equation}%
which has a transition between quadratic and linear behaviors when $\gamma
t\ll 1$ and $\gamma t\gg 1$ respectively. This expression for $\gamma _{t}$
indicates that a colored Gaussian noise approximation (with an exponential
noise correlation) describes the behavior of the standard LE and of $\mathrm{%
E}_{\pm \mp }(t,\tau )$~(Appendix \ref{AppBoson}).

However, under the same assumptions one also finds that $\phi
_{t}=(A/2)(1-e^{-\gamma t}),$ which implies%
\begin{equation}
\Phi _{t,\tau }=\frac{A}{2}(1-e^{-\gamma t})(1-e^{-\gamma \tau }).
\end{equation}%
This phase is present even though we are considering the high temperature
regime of a harmonic system, and it is a signature of the quantum character
of the operators governing the dynamics.

In Fig.~2 we plot the differences $|$\textrm{$E$}$_{++}(t,\tau )-\mathrm{E}%
_{--}(t,\tau )|$ and $|\mathrm{E}_{+-}(t,\tau )-\mathrm{E}(\tau )\mathrm{E}%
^{\ast }(t)|.$ The inset shows the LE decay. For the parameters chosen, this
decay starts to approach an exponential behavior, $\mathrm{E}(\tau )=\exp
[-\gamma _{\tau }]\simeq \exp [-A/(\beta \hbar \gamma )(\gamma \tau -1)]$ $%
(\gamma \tau >1)$. In spite of this property, fingerprints of the underlying
quantum dynamics (left panel) and deviations from a white noise limit (right
panel) are still observed.

The phase $\Phi _{t,\tau },$ which measures departures from classicality,
only vanishes if one takes an additional limit \mbox{$\gamma \to
\infty$}, $\beta \to 0,$ and $A\to 0$. This limit is taken with $\beta \hbar
\gamma \ll 1$ and with $A\propto \beta$, so that $A\to0$ while $A/\beta
\hbar =\alpha$ remains finite. This leads to an exact exponential decay of
the LE, $\mathrm{E}(\tau )=\exp [-\alpha \tau ],$ and $\Phi _{t,\tau }=0.$
Then a classical noise representation is valid, with delta-correlated noise.
Indeed, in this limit one gets $|$\textrm{$E$}$_{++}(t,\tau )-\mathrm{E}%
_{--}(t,\tau )|\to 0$ and $|\mathrm{E}_{+-}(t,\tau )-\mathrm{E}(\tau )%
\mathrm{E}^{\ast }(t)|\to 0$. 
\begin{figure}[t]
\includegraphics[bb=47 872 726
1140,angle=0,width=8.5cm]{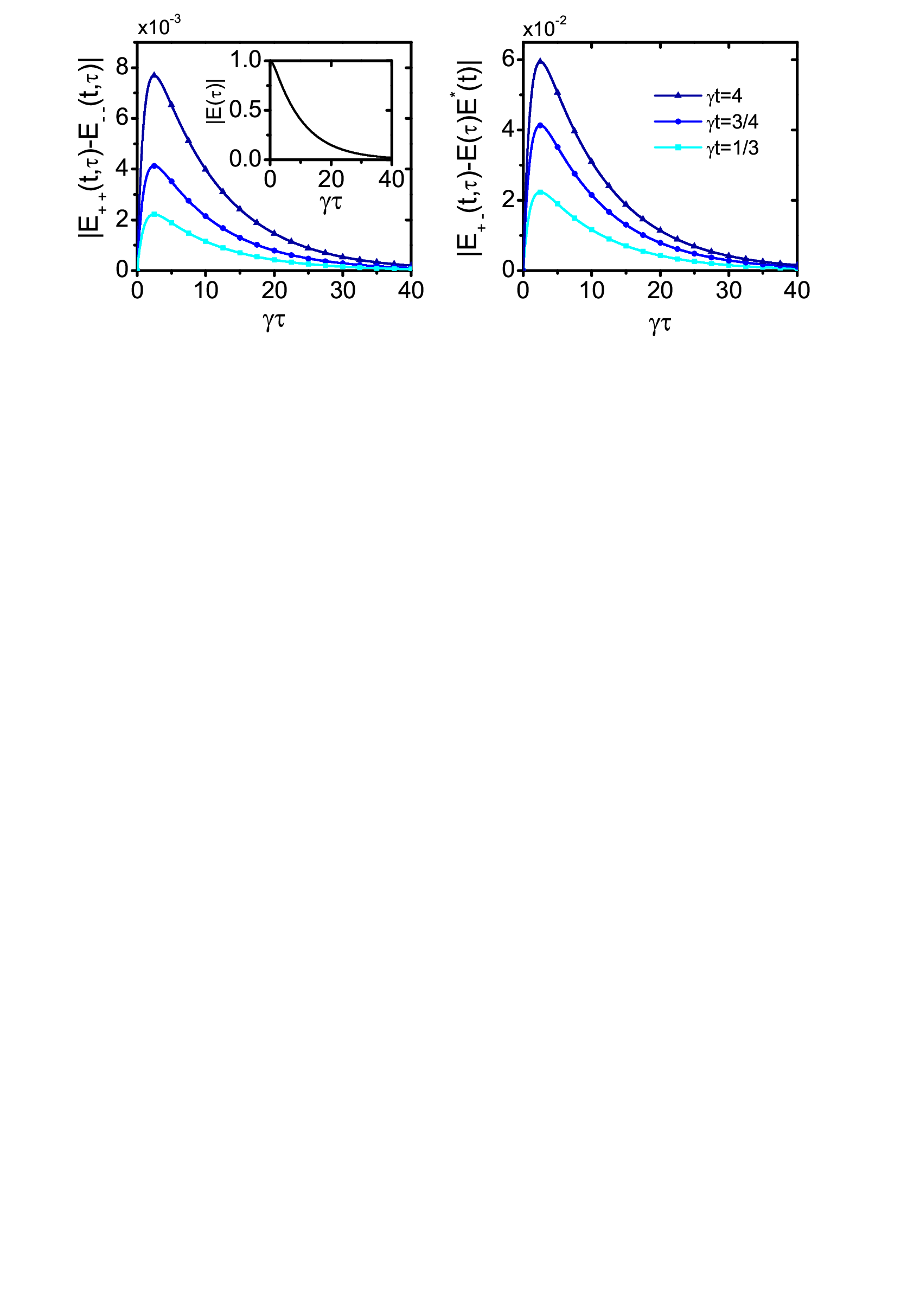}
\caption{Non-classicality quantifier $|\mathrm{E}_{++}(t,\protect\tau )-%
\mathrm{E}_{--}(t,\protect\tau )|$ (left panel) and departure from a white
noise limit $|\mathrm{E}_{+-}(t,\protect\tau )-\mathrm{E}(\protect\tau )%
\mathrm{E}^{\ast }(t)|$ (right panel) for the bosonic dynamics Eq.~(\protect
\ref{bosonicHI}) with an Ohmic spectral density (see text). The parameters
are $\protect\beta \hbar \protect\gamma =10^{-1}$ and $A=10^{-2}.$ The inset
shows the standard LE, which displays an approximately exponential decay.}
\end{figure}

\section{Conclusions}

We have presented a set of generalizations of the well established Loschmidt
echo. In contrast to the latter, which is given by a state overlap after the
application of two time propagators, the generalized echoes involve
different combinations of four time propagators. These correspond to
step-wise evolutions generated by two different Hamiltonians, combining two
steps of forward evolution followed by two steps of backward evolution. This
simple extension of the original echo admits an analysis and implementation
in forms that are very similar to the standard case.

The advantage of this family of generalized echoes is the ability to
directly identify signatures of quantum behavior in the time evolution. More
precisely, differences between echoes can be directly associated with
non-vanishing commutators of time propagators. Additionally, one can also
use combinations of echoes to identify departures from different models of
classical noise, such as white noise or noise with infinite time
correlations. These features were supported by explicit results in
paradigmatic cases such as critical spin dynamics and harmonic oscillators
in a high temperature limit. Thus, the generalized echoes provide a simple
and useful tool for the characterization and quantification of quantum
features in the dynamics of interest.

\section*{Acknowledgments}

A.A.B.\ thanks finantial support from CONICET (Argentina). C.C. acknowledges
funding from CSIC and PEDECIBA (Uruguay).

\appendix

\section{Classical noise with Gaussian statistics \label{GaussNoise}}

In the noise approach, the standard LE reads,%
\begin{equation}
\mathrm{E}(\tau )=\langle \exp [-i\textstyle\int_{0}^{\tau }\xi (t^{\prime
})dt^{\prime }]\rangle ,
\end{equation}%
where $\xi (t)$ is a classical scalar noise and $\langle \cdots \rangle $\
denotes an average over realizations. The GLEs are 
\begin{subequations}
\begin{eqnarray}
\mathrm{E}_{++}(t,\tau ) &=&\langle \exp [-i\textstyle\int_{t}^{t+\tau }\xi
(t^{\prime })dt^{\prime }]\rangle , \\
\mathrm{E}_{--}(t,\tau ) &=&\langle \exp [-i\textstyle\int_{t}^{t+\tau }\xi
(t^{\prime })dt^{\prime }]\rangle , \\
\mathrm{E}_{+-}(t,\tau ) &=&\langle \exp [-i\textstyle\int_{t}^{t+\tau }\xi
(t^{\prime })dt^{\prime }+i\textstyle\int_{0}^{t}\xi (t^{\prime })dt^{\prime
}]\rangle ,\ \ \ \ \ \ \  \\
\mathrm{E}_{-+}(t,\tau ) &=&\langle \exp [-i\textstyle\int_{0}^{t+\tau }\xi
(t^{\prime })dt^{\prime }]\rangle .
\end{eqnarray}%
It is always fulfilled that\ $\mathrm{E}_{-+}(t,\tau )=\mathrm{E}(t+\tau ).$
For stationary noises, the only GLE that cannot be written in terms of $%
\mathrm{E}(\tau )$ is the echo $\mathrm{E}_{+-}(t,\tau ).$ In fact, under
the stationary condition $\mathrm{E}_{++}(t,\tau )=\mathrm{E}_{--}(t,\tau )=%
\mathrm{E}(\tau ).$

The previous expressions for the LE and its generalized versions can be
expressed in terms of the characteristic functional of the noise. For an
arbitrary function $k(t)$ it is defined as~\cite{kampen} 
\end{subequations}
\begin{equation}
G[k]\equiv \left\langle \exp \left[ i\int_{0}^{\infty }k(t^{\prime })\xi
(t^{\prime })dt^{\prime }\right] \right\rangle .
\end{equation}%
Consequently, it is possible to write $\mathrm{E}(\tau )=G[k_{E}],$ where $%
k_{E}(t^{\prime })=-\theta (\tau -t^{\prime })$ with $\theta (x)$ the step
function. Similarly, $\mathrm{E}_{+-}(t,\tau )=G[k_{E_{+-}}],$ where $%
k_{E_{+-}}(t^{\prime })=-\theta (t^{\prime }-t)\theta (t+\tau -t^{\prime
})+\theta (t-t^{\prime }).$

The characteristic functional of a \textit{Gaussian noise} with null mean, $%
\langle \xi (t)\rangle =0,$ is~\cite{kampen}%
\begin{equation}
G[k]=\exp \!\left[ -\frac{1}{2}\int_{0}^{\infty
}\!\!\!dt_{2}\int_{0}^{\infty }\!\!\!dt_{1}k(t_{2})k(t_{1})\langle \langle
\xi (t_{2})\xi (t_{1})\rangle \rangle \right] .
\end{equation}%
Here, $\langle \langle \xi (t_{2})\xi (t_{1})\rangle \rangle $ is the noise
correlation. For stationary noise, we define the noise correlation function
as $f(|t_{2}-t_{1}|)=\langle \langle \xi (t_{2})\xi (t_{1})\rangle \rangle .$
Using this condition, with the proper function $k(t)$ as defined previously,
the LE reads%
\begin{equation}
\mathrm{E}(\tau )=\exp (-\gamma _{\tau }),
\end{equation}%
with the decay determined by the dimensionless exponent%
\begin{equation}
\gamma _{\tau }=\int_{0}^{\tau }dt_{2}\int_{0}^{\tau }dt_{1}f(|t_{2}-t_{1}|).
\label{GamaLess}
\end{equation}%
Notice that $(d^{2}/d^{2}\tau )\gamma _{\tau }=2f(|\tau |)$ recovers the
stationary noise correlation.

Clearly, from the above expressions for stationary noise we also have%
\begin{equation}
\mathrm{E}_{\pm \pm }(t,\tau )=\exp (-\gamma _{\tau }).
\end{equation}%
Finally, after some rearrangement of the corresponding integrals, for the
remaining GLEs one obtains%
\begin{equation}
\mathrm{E}_{\pm \mp }(t,\tau )=\exp [-(\gamma _{\tau }+\gamma _{t}\pm \Gamma
_{t,\tau })],
\end{equation}%
where%
\begin{equation}
\Gamma _{t,\tau }=\gamma _{t}+\gamma _{\tau }-\gamma _{t+\tau }.
\end{equation}

As a particular example, we consider an exponential correlation function%
\begin{equation}
f(|t_{2}-t_{1}|)=2g^{2}\exp (-|t_{2}-t_{1}|/\tau _{c}),  \label{ExpoCorre}
\end{equation}%
where $g^{2}$ scales the noise amplitude and $\tau _{c}$ is the
time-correlation. Then, from Eq.~(\ref{GamaLess}) it follows%
\begin{equation}
\gamma _{\tau }=4 g^{2}\tau _{c}^{2}\left[ \frac{\tau }{\tau _{c}}%
-(1-e^{-\tau /\tau _{c}})\right] .  \label{GamaNoise}
\end{equation}

\section{Quantum Ising model with a transverse field \label{Ising}}

Here we calculate the echoes associated to a spin chain with different
transverse fields. The Hamiltonian corresponding to each option, $\lambda
\longrightarrow \lambda ^{(\pm )},$ is%
\begin{equation}
H(\lambda )=-J\sum_{j=1}^{N}\lambda \sigma _{j}^{(x)}+\sigma
_{j}^{(z)}\sigma _{j+1}^{(z)}.  \label{Halfa}
\end{equation}%
$N$ is the number of spins, $\sigma _{j}^{(x)}$ and $\sigma _{j}^{(z)}$ are
the Pauli operators in the $x$- and $z$-directions respectively. In a first
step, this Hamiltonian is diagonalized. In a second step, the technique that
allows to calculate the GLEs in an exact analytical way is developed. In the
third step, the explicit expressions for the GLEs are provided.

\subsection{Hamiltonian diagonalization}

Following Ref.~\cite{Dziarmaga_2005} periodic boundary conditions are
considered such that $\vec{\sigma}_{N+1}=\vec{\sigma}_{1}.$ In the following
calculations the echo index is drop from $\lambda $ and $J\to 1.$

\textit{Mapping of spin chain to fermion chain: }After a Jordan-Wigner
transformation~\cite{Dziarmaga_2005} , the spin system can be mapped onto a
system of fermions, with Hamiltonian%
\begin{equation}
H(\lambda )=P_{+}\mathcal{H}_{+}P_{+}+P_{-}\mathcal{H}_{-}P_{-},
\label{eq:Ham splitting}
\end{equation}%
where $P_{\pm }$ project onto subspaces with defined parity, as $P_{\pm }=%
\frac{1}{2}\left[ 1\pm \Pi \right] ,$ with%
\begin{equation}
\Pi =\prod_{j=1}^{N}\sigma _{j}^{(x)}=\prod_{j=1}^{N}(1-2c_{j}^{\dagger
}c_{j}).
\end{equation}%
For notational convenience the subspaces are indexed with $\pm $ (they do
not have any relation with the GLEs notation). The Hamiltonians in fermionic
version are written as%
\begin{equation}
\mathcal{H}_{\pm }=\sum_{j=1}^{N}\Big[2\lambda c_{j}^{\dagger
}c_{j}-c_{j}^{\dagger }c_{j+1}-c_{j+1}^{\dagger
}c_{j}-c_{j+1}c_{j}-c_{j}^{\dagger }c_{j+1}^{\dagger }-\lambda \Big],
\end{equation}%
which looks exactly the same for both cases, but the boundary conditions are
different: for the \textquotedblleft even\textquotedblright\ subspace
corresponding to $P_{+}$ they are antiperiodic, i.e. $c_{N+1}=-c_{1}$, while
they are periodic for the \textquotedblleft odd\textquotedblright\ subspace,
i.e. $c_{N+1}=c_{N}.$ Notice that since the Hamiltonians preserve parity,
the use in~(\ref{eq:Ham splitting}) of two projection operators, one at each
side of the Hamiltonian, is not really necessary but is chosen for clarity.

Both Hamiltonians $\mathcal{H}_{\pm }$ can be diagonalized by means of a
Fourier transformation followed by a Bogoliubov transformation which
factorizes over the spaces with different pseudomomentum $k$. But because of
the different boundary conditions, the Fourier transformation is defined
differently according to the parity. Here, only the even case is solved
because the parity is preserved in this problem and the ground state of the
Ising chain always corresponds to an even state. In any case, the boundary
conditions become irrelevant for large $N.$

\textit{Diagonalization of the even subspace: }In order to diagonalized $%
\mathcal{H}_{+}$, one defines a Fourier transformation in the form%
\begin{equation}
c_{j}=\frac{e^{-i\pi /4}}{\sqrt{N}}\sum_{k\in K_{+}}c_{k}e^{ikj},
\end{equation}%
with values of pseudomomentum given by $K_{\pm }=\{k=\pm (\pi /N)(2n-1),$
with $n=1,\ldots ,N/2\},$ where for definiteness $N$ is taken to be even.
After this transformation the Hamiltonian takes the form%
\begin{equation}
\mathcal{H}_{+}\!=\!\sum_{k\in K_{+}}\Big\{2[\lambda -\cos
(k)]c_{k}^{\dagger }c_{k}+\sin (k)(c_{k}^{\dagger }c_{-k}^{\dagger
}+c_{-k}c_{k})-\lambda \Big\},
\end{equation}%
where it is important to note that all the terms preserve pseudomomentum,
and therefore the diagonalization can be done in blocks $\pm k.$ The last
step is a Bogoliubov transformation defined as%
\begin{align}
c_{k}& =\gamma _{k}\cos (\phi _{k}/2)-\gamma _{-k}^{\dagger }\sin (\phi
_{k}/2),  \label{eq:ck in terms of gamma} \\
c_{-k}& =\gamma _{-k}\cos (\phi _{k}/2)+\gamma _{k}^{\dagger }\sin (\phi
_{k}/2).  \label{eq:c-k in terms of gamma}
\end{align}%
It is easy to check that these new operators also satisfy the corresponding
anticommutation relations. Furthermore,%
\begin{equation}
\cos (\phi _{k})=\frac{2[\lambda -\cos (k)]}{\epsilon _{k}},\ \ \ \ \ \ \ \
\ \sin (\phi _{k})=\frac{2\sin (k)}{\epsilon _{k}},  \label{Phik}
\end{equation}%
where%
\begin{equation}
\epsilon _{k}=2\sqrt{\sin ^{2}(k)+[\lambda -\cos (k)]^{2}}=2\sqrt{1+\lambda
^{2}-2\lambda \cos (k)}.  \label{Energyk}
\end{equation}%
In this way the final diagonal version is found,%
\begin{equation}
\mathcal{H}_{+}=\sum_{k\in K_{+}}\epsilon _{k}\left( \gamma _{k}^{\dagger
}\gamma _{k}-\frac{1}{2}\right) \,.  \label{HPlus}
\end{equation}%
With this definition $\epsilon _{k}$ is the cost of creating a quasiparticle
with pseudomomentum $k$. Note that $\epsilon _{k}=\epsilon _{-k}>0$ and that
the total spectrum is symmetric with respect to the zero of energy. This
makes senses because it is possible to apply spin rotations such that $%
H(\lambda )$ is transformed into $-H(\lambda ).$

\subsection{Calculation of the echoes}

Given that the Hamiltonians does not depends on time [Eq. (\ref{Halfa})],
for simplifying the notation, the GLEs $\mathrm{E}_{\tilde{s}s}(t,\tau
)=\langle \mathcal{B}|\mathbb{G}_{t,0}^{(\tilde{s})\dag }\mathbb{G}_{t+\tau
,t}^{(-)\dag }\mathbb{G}_{t+\tau ,t}^{(+)}\mathbb{G}_{t,0}^{(s)}|\mathcal{B}%
\rangle $ are expressed as $\mathrm{E}_{\tilde{s}s}(t,\tau )\to \mathrm{E}%
^{abcd}(t,\tau ),$\ where%
\begin{equation}
\mathrm{E}^{abcd}(t,\tau )=\langle \mathcal{B}|\mathbb{G}_{t}^{(a)\dagger }%
\mathbb{G}_{\tau }^{(b)\dagger }\mathbb{G}_{\tau }^{(c)}\mathbb{G}_{t}^{(d)}|%
\mathcal{B}\rangle .  \label{eq:echoes}
\end{equation}%
Here, $|\mathcal{B}\rangle $ is the initial state. The supraindexes $a,b,c,d$
label the different time-independent Hamiltonians [Eq.~(\ref{Halfa}) with
different values of $\lambda =\lambda ^{(\pm )}$] governing the evolution in
each time step, and the times indicates the duration of this evolution step.
The full calculation is then a step-wise evolution. With this notation, the
standard LE and the GLEs can be calculated in the same way.

Since the splitting in terms of quasimomentum blocks $\pm k$ holds for any
value of $\lambda $ [see Eq.~(\ref{HPlus})], the calculation of the echoes~(%
\ref{eq:echoes}) can always be factorized,%
\begin{equation}
\mathrm{E}^{abcd}(t,\tau )=\prod_{k>0}\mathrm{E}_{\pm k}^{abcd}(t,\tau ),
\label{eq:echoes_factorized}
\end{equation}%
that is, the full echo is a product of equivalent echoes but defined only on
the $\pm k$ Fourier modes,%
\begin{equation}
\mathrm{E}_{\pm k}^{abcd}(t,\tau )=[\langle \mathcal{B}|\mathbb{G}%
_{t}^{(a)\dagger }\mathbb{G}_{\tau }^{(b)\dagger }\mathbb{G}_{\tau }^{(c)}%
\mathbb{G}_{t}^{(d)}|\mathcal{B}\rangle ]_{\pm k}.  \label{eq:echoes_factors}
\end{equation}

In the following two calculus steps, for shortening the notation the two
Hamiltonians [Eq.~(\ref{Halfa})] associated to the two possible values of
the transverse field are denoted \textit{without} tilde for $%
H^{(-)}=H(\lambda ^{(-)})$ and \textit{with} tilde for $H^{(+)}=H(\lambda
^{(+)}),$ and equivalently for the quasiparticle operators in Eqs.~(\ref%
{eq:ck in terms of gamma}) and (\ref{eq:c-k in terms of gamma}). Making use
of the relations (\ref{eq:ck in terms of gamma}) and (\ref{eq:c-k in terms
of gamma}), relating the $c$ with the $\gamma $ operators, and the analogous
relations for $\tilde{\gamma}$, one finds that%
\begin{align}
\tilde{\gamma}_{k}& =\gamma _{k}\cos (\Delta _{k}/2)+\gamma _{-k}^{\dagger
}\sin (\Delta _{k}/2), \\
\tilde{\gamma}_{-k}& =\gamma _{-k}\cos (\Delta _{k}/2)-\gamma _{k}^{\dagger
}\sin (\Delta _{k}/2),
\end{align}%
where $\Delta _{k}=\tilde{\phi}_{k}-\phi _{k}.$ Furthermore, because the
parity in quasiparticle number is preserved, and because the initial state $|%
\mathcal{B}\rangle $ is taken as the ground state of $H^{(-)},$ we only need
to use the following relations 
\begin{subequations}
\label{Rotor}
\begin{align}
\widetilde{|{0,0}\rangle }& =\cos (\Delta _{k}/2)|{0,0}\rangle -\sin (\Delta
_{k}/2)|{1,1}\rangle , \\
\widetilde{|{1,1}\rangle }& =\sin (\Delta _{k}/2)|{0,0}\rangle +\cos (\Delta
_{k}/2)|{1,1}\rangle ,
\end{align}%
where $|{0,0}\rangle $ is the state annihilated by $\gamma _{\pm k}$ while $|%
{1,1}\rangle =\gamma _{k}^{\dagger }\gamma _{-k}^{\dagger }|{0,0}\rangle $,
and equivalently with the tilde.

Using all this, the echoes can be written entirely in terms of the
change-of-basis matrix $C$ defined from Eq.~(\ref{Rotor}) as 
\end{subequations}
\begin{equation}
C=%
\begin{pmatrix}
\cos (\Delta _{k}/2) & -\sin (\Delta _{k}/2) \\ 
\sin (\Delta _{k}/2) & \cos (\Delta _{k}/2)%
\end{pmatrix}%
,
\end{equation}%
where 
\begin{equation}
\Delta _{k}=\phi _{k}^{(+)}-\phi _{k}^{(-)},  \label{DeltaPhi}
\end{equation}%
and the diagonal matrices that provide the phases accumulated by an
evolution step of the desired time, 
\begin{equation}
D_{t}^{(\pm )}=%
\begin{pmatrix}
e^{i\epsilon _{k}^{(\pm )}t} & 0 \\ 
0 & e^{-i\epsilon _{k}^{(\pm )}t}%
\end{pmatrix}%
,  \label{Dpm}
\end{equation}%
where $(+)$ and $(-)$ are the explicit GLEs indexes. In Eq.~(\ref{DeltaPhi}) 
$\phi _{k}^{(\pm )}$ are defined from Eq.~(\ref{Phik}) with $\lambda \to
\lambda ^{(\pm )},$ while in Eq.~(\ref{Dpm}) $\epsilon _{k}^{(\pm )}$ are
defined from Eq.~(\ref{Energyk}) with $\lambda \to \lambda ^{(\pm )}.$ The
matrices $C,$ $D_{t}^{(\pm )}$ depend on $k$, but we leave this dependence
implicit since it is clear enough.

In the end, one then obtains,%
\begin{equation}
\mathrm{E}_{\pm k}^{abcd}(t,\tau )=\langle {e}_{0}|M_{t}^{(a)\dagger
}M_{\tau }^{(b)\dagger }M_{\tau }^{(c)}M_{t}^{(d)}|e_{0}\rangle ,
\label{Echo_Mk}
\end{equation}%
where $|e_{0}\rangle \equiv (1,0)^{\intercal }$ takes into account the
initial condition. Each two-dimensional matrix is defined as%
\begin{equation}
M_{t}^{(a)}=D_{t}^{(-)}\ \ \ \ \ \ if\ \ a=(-),
\end{equation}%
and%
\begin{equation}
M_{t}^{(a)}=C^{\intercal }D_{t}^{(+)}C\ \ \ \ \ \ if\ \ a=(+).
\end{equation}%
Similarly for $b,$ $c,$ $d,$ and $t\to\tau .$ In Eq.~(\ref{Echo_Mk}), the $k$%
-dependence left implicit in the matrices to avoid too cumbersome notation.

\subsection{Explicit expressions for GLEs}

From Eqs.~(\ref{eq:echoes_factorized}) and~(\ref{Echo_Mk}), taking $t=0,$
the standard LE can be written as%
\begin{equation}
\mathrm{E}(\tau )=\prod_{k}\mathrm{E}_{k}(\tau ),
\end{equation}%
where each $k$-component reads%
\begin{equation}
\mathrm{E}_{k}(\tau )\!=\!e^{-i\tau \varepsilon _{k}^{(-)}}\![e^{+i\tau
\varepsilon _{k}^{(+)}}\!\cos ^{2}(\Delta _{k}/2)+e^{-i\tau \varepsilon
_{k}^{(+)}}\!\sin ^{2}(\Delta _{k}/2)].  \label{Echo_k}
\end{equation}%
Here, $\varepsilon _{k}^{(\pm )}$ are the energy of each independent Fourier
mode, Eq.~(\ref{Energyk}) with $\lambda \to \lambda ^{(\pm )}.$ Furthermore,
in $\Delta _{k}=\phi _{k}^{(+)}-\phi _{k}^{(-)}$ each contribution $\phi
_{k}^{(\pm )}$ is defined from Eq.~(\ref{Phik}) with $\lambda \to \lambda
^{(\pm )}.$

On the other hand, from a similar and explicit calculus based on Eqs.~(\ref%
{eq:echoes_factorized}) and the matrix product~(\ref{Echo_Mk}), we get%
\begin{equation}
\mathrm{E}_{\tilde{s}s}(t,\tau )=\prod_{k}\mathrm{E}_{k}^{\tilde{s}s}(t,\tau
),
\end{equation}%
where 
\begin{subequations}
\begin{eqnarray}
\mathrm{E}_{k}^{++}(t,\tau ) &=&\mathrm{E}_{k}(\tau )+\mathrm{C}%
_{k}^{++}(t,\tau ), \\
\mathrm{E}_{k}^{--}(t,\tau ) &=&\mathrm{E}_{k}(\tau ), \\
\mathrm{E}_{k}^{+-}(t,\tau ) &=&\mathrm{E}_{k}(\tau -t)+\mathrm{C}%
_{k}^{+-}(t,\tau ), \\
\mathrm{E}_{k}^{-+}(t,\tau ) &=&\mathrm{E}_{k}(\tau +t).
\end{eqnarray}%
The extra contributions are 
\end{subequations}
\begin{eqnarray}
\mathrm{C}_{k}^{++}(t,\tau ) &=&2i\sin (\varepsilon _{k}^{(+)}t)\sin
[\varepsilon _{k}^{(+)}(\tau +t)]  \notag \\
&&\times \sin (\varepsilon _{k}^{(-)}\tau )\sin ^{2}(\Delta _{k}),
\end{eqnarray}%
and similarly%
\begin{eqnarray}
\mathrm{C}_{k}^{+-}(t,\tau ) &=&2i\sin (\varepsilon
_{k}^{(+)}t)[e^{i\varepsilon _{k}^{(-)}t}\sin (\varepsilon _{k}^{(+)}\tau )]
\notag \\
&&\times \sin (\varepsilon _{k}^{(-)}\tau )\sin ^{2}(\Delta _{k}).
\end{eqnarray}%
These last two contributions gives departures with respect to a random
frequency representation.

\section{Quantum harmonic oscillators \label{AppBoson}}

Here we consider a set of non-interacting quantum harmonic oscillators $%
H=\sum_{k}\omega _{k}b_{k}^{\dag }b_{k}$ with commutation relations $%
[b_{k},b_{k^{\prime }}^{\dagger }]=\delta _{kk^{\prime }}.$ The dynamics is
symmetrically modified by a linear force operator. In an interaction
representation the propagators that define the GLEs are $\mathbb{G}%
_{t,t_{0}}^{(\pm )}=\exp [-i\int_{t_{0}}^{t}dt^{\prime }H^{(\pm )}(t^{\prime
})]$ where%
\begin{equation}
H^{\pm }(t)=\pm \sum_{k}(g_{k}e^{+i\omega _{k}t}b_{k}^{\dag }+g_{k}^{\ast
}e^{-i\omega _{k}t}b_{k}).
\end{equation}

\subsection{Propagators}

From standard tools of open quantum system theory$~$\cite{Breuer_2002}, the
propagator $\mathbb{G}_{t,t_{0}}^{(\pm )}$ can be expressed as%
\begin{equation}
\mathbb{G}_{t,t_{0}}^{(\pm )}=\prod_{k}\mathbb{D}_{k}[\pm \alpha
_{t,t_{0}}^{(k)}]\exp (i\varphi _{t,t_{0}}).  \label{PropasPlusMinus}
\end{equation}%
The time-dependent coefficients are%
\begin{equation}
\alpha _{t,t_{0}}^{(k)}=g_{k}e^{+i\omega _{k}t_{0}}f_{k}(t-t_{0}),
\end{equation}%
where the auxiliary functions read%
\begin{equation}
f_{k}(t)=\frac{1-e^{+i\omega _{k}t}}{\omega _{k}},\ \ \ \ \ \ \ f_{k}^{\ast
}(t)=-f_{k}(t)e^{-i\omega _{k}t}.  \label{efesK}
\end{equation}%
The global phase in Eq.$~$(\ref{PropasPlusMinus}) is $\varphi
_{t,t_{0}}=\int\nolimits_{t_{0}}^{t}dt^{\prime
}\int\nolimits_{t_{0}}^{t^{\prime }}dt^{\prime \prime
}\sum\nolimits_{k}|g_{k}|^{2}\sin [\omega _{k}(t^{\prime }-t^{\prime \prime
})],$ which in this example is irrelevant in the calculation of the LE and
the GLEs. Finally, in Eq.$~$(\ref{PropasPlusMinus}) the displacement
operators associated to each bosonic mode are%
\begin{equation}
\mathbb{D}_{k}[\alpha ]\equiv \exp (\alpha b_{k}^{\dag }-\alpha ^{\ast
}b_{k}).
\end{equation}%
They satisfy $\mathbb{D}_{k}^{\dag }[\alpha ]=\mathbb{D}_{k}[-\alpha ].$

\subsection{Calculation of the echoes}

The product of two displacement operators with different coefficients satisfy%
$~$\cite{Breuer_2002}%
\begin{equation}
\mathbb{D}_{k}[\alpha ]\mathbb{D}_{k}[\beta ]=\mathbb{D}_{k}[\alpha +\beta
]\exp [(\alpha \beta ^{\ast }-\alpha ^{\ast }\beta )/2].
\end{equation}%
This rule allows us to solve the product of different propagators appearing
in the definition of the GLE, that is,%
\begin{equation}
\mathbb{G}_{t,0}^{(\tilde{s})\dag }\mathbb{G}_{t+\tau ,t}^{(-)\dag }\mathbb{G%
}_{t+\tau ,t}^{(+)}\mathbb{G}_{t,0}^{(s)}=\prod\nolimits_{k}\mathbb{D}%
_{k}[\alpha _{t,\tau ,\tilde{s},s}^{(k)}]\exp (\phi _{t,\tau ,\tilde{s}%
,s}^{(k)}).  \label{PropaProduct}
\end{equation}%
Here, the total displacement amplitude $\alpha _{t,\tau ,\tilde{s},s}^{(k)}$%
\ and phase $\phi _{t,\tau ,\tilde{s},s}^{(k)}$ depend on each GLE case,
which are parametrized by $s$ and $\tilde{s}.$

The LE and its generalization can be explicitly calculated after specifying
the initial state, which allows to take the expectation value of the
displacement operators defined by Eq.$~$(\ref{PropaProduct}). Considering a
thermal state, the mean value of a displacement operator reads$~$\cite%
{Breuer_2002}%
\begin{equation}
\langle \mathbb{D}_{k}[\alpha ]\rangle _{\mathrm{th}}=\exp [-|\alpha
|^{2}\langle \{b_{k},b_{k}^{\dag }\}_{+}\rangle _{\mathrm{th}}/2],
\label{ThermalAverage}
\end{equation}%
where the brackets $\langle \mathbb{\cdots }\rangle _{\mathrm{th}}$ denotes
expectation value with respect to a (bosonic) thermal state. On the other
hand, from the bosonic commutation relations $\langle \{b_{k},b_{k}^{\dag
}\}_{+}\rangle _{\mathrm{th}}=1+2\langle b_{k}^{\dag }b_{k}\rangle _{\mathrm{%
th}},$ it follows%
\begin{equation}
\langle \{b_{k},b_{k}^{\dag }\}_{+}\rangle _{\mathrm{th}}=1+2\left( \frac{%
e^{-\beta \hbar \omega _{k}}}{1-e^{-\beta \hbar \omega _{k}}}\right) =\coth
(\beta \hbar \omega _{k}/2).  \label{NThermal}
\end{equation}%
Here, $\beta =1/k_{B}T$ where $T$ is the environmental temperature and $k_{B}
$ is the Boltzmann constant.

\subsection{Explicit expressions for GLEs}

The results expressed by Eqs. (\ref{PropaProduct}) and (\ref{ThermalAverage}%
) allow the GLEs be calculated in an exact analytical way. We get%
\begin{equation}
\mathrm{E}(\tau )=\exp (-\gamma _{\tau }),
\end{equation}%
where the dimensionless rate is%
\begin{equation}
\gamma _{t}=2\sum_{k}|g_{k}|^{2}|f_{k}(t)|^{2}\coth (\beta \hbar \omega
_{k}/2).
\end{equation}%
The GLEs are%
\begin{equation}
\mathrm{E}_{\pm \pm }(t,\tau )=\exp (-\gamma _{\tau }\pm i\Phi _{t,\tau }),
\end{equation}%
where%
\begin{equation}
i\Phi _{t,\tau }=2\sum_{k}|g_{k}|^{2}[f_{k}(t)f_{k}(\tau )-f_{k}^{\ast
}(t)f_{k}^{\ast }(\tau )].
\end{equation}%
After some rearrangement of the corresponding integrals, it is also possible
to obtain%
\begin{equation}
\mathrm{E}_{\pm \mp }(t,\tau )=\exp [-(\gamma _{\tau }+\gamma _{t}\pm \Gamma
_{t,\tau })],
\end{equation}%
where%
\begin{equation}
\Gamma _{t,\tau }=2\sum_{k}|g_{k}|^{2}[f_{k}(t)f_{k}(\tau )+f_{k}^{\ast
}(t)f_{k}^{\ast }(\tau )]\coth (\beta \hbar \omega _{k}/2),
\end{equation}%
Using the explicit definition of the functions $\{f_{k}(t)\}$ [Eq.~(\ref%
{efesK})] one finds the final expressions in the manuscript, that is%
\begin{equation}
\gamma _{t}=4\sum_{k}\frac{|g_{k}|^{2}}{\omega _{k}^{2}}[1-\cos (\omega
_{k}t)]\coth (\beta \hbar \omega _{k}/2),  \label{GammatimeApp}
\end{equation}%
where in addition $\Gamma _{t,\tau }=\gamma _{t}+\gamma _{\tau }-\gamma
_{t+\tau }.$ Furthermore, $\Phi _{t,\tau }=\phi _{t}+\phi _{\tau }-\phi
_{t+\tau },$ where%
\begin{equation}
\phi _{t}=4\sum_{k}\frac{|g_{k}|^{2}}{\omega _{k}^{2}}\sin (\omega _{k}t).
\label{PhasetimeApp}
\end{equation}

\subsubsection*{Ohmic spectral density}

In the microscopic Hamiltonian description, the functions $\gamma _{t}$ and $%
\phi _{t}$ can be explicitly evaluated after specifying the spectral
properties. A continuous limit is considered, such that $\sum%
\nolimits_{k}4|g_{k}|^{2}\to \int_{0}^{\infty }d\omega J(\omega ).$ The
spectral density reads $J(\omega )=4|g_{\omega }|^{2}D(\omega ),$ where $%
D(\omega )$ is the density of modes at frequency $\omega .$ We consider a
ohmic case$~$\cite{Breuer_2002} with a Lorentzian high frequency cutoff,%
\begin{equation}
J(\omega )=A\omega \frac{\gamma ^{2}/\pi }{(\omega ^{2}+\gamma ^{2})}.
\label{OhmicSpectral}
\end{equation}%
Below, we obtain the rate $\gamma _{t}$ \ and phase $\phi _{t}.$

Under the previous assumptions, from Eq.$~$(\ref{GammatimeApp}) it follows
that%
\begin{equation}
\gamma _{t}=\int_{0}^{\infty }d\omega J(\omega )\frac{[1-\cos (\omega t)]}{%
\omega ^{2}}\coth (\beta \hbar \omega /2).  \label{GamaIntegral}
\end{equation}%
This integral expression can be calculated in an exact analytical way. Using
that$~$\cite{Breuer_2002}%
\begin{equation}
\coth (\beta \hbar \omega /2)=\frac{2}{\beta \hbar }\sum_{n=-\infty
}^{+\infty }\frac{\omega }{\omega ^{2}+\nu _{n}^{2}},
\end{equation}%
where the Matsubara frequencies are $\nu _{n}=2\pi n/(\beta \hbar ),$ the
method of residues leads to the exact expression%
\begin{eqnarray}
\gamma _{t} &=&\frac{A}{\beta \hbar \gamma }(\gamma t)-\frac{A}{2}%
(1-e^{-\gamma t})\cot (\beta \hbar \gamma /2)  \notag \\
&&-\frac{2A}{\beta \hbar \gamma }\sum_{n=1}^{\infty }\frac{(1-e^{-\mu
_{n}\gamma t})}{\mu _{n}(\mu _{n}^{2}-1)},  \label{GamaExacta}
\end{eqnarray}%
with $\mu _{n}\equiv \nu _{n}/\gamma =2\pi n/(\beta \hbar \gamma ).$

In a \textit{high-temperature limit} $(\beta \hbar \gamma \ll 1),$ the
expression$~$(\ref{GamaExacta}) can be approximated as%
\begin{equation}
\gamma _{t}\simeq \frac{A}{\beta \hbar \gamma }[\gamma t-(1-e^{-\gamma t})].
\label{GamaHighT}
\end{equation}%
This structure also emerges from \textit{Gaussian noise approximation}.
Considering the stationary noise correlation defined by Eq.$~$(\ref%
{ExpoCorre}), under the associations $\tau _{c}^{-1}\leftrightarrow \gamma $
and $4(\tau _{c}g)^{2}\leftrightarrow A/\beta \hbar \gamma ,$ the decay rate 
$\gamma _{t}$ defined by Eqs.$~$(\ref{GamaNoise}) and$~$(\ref{GamaHighT})
are indistinguishable.

For intermediate temperature, all contributions proportional to the
Matsubara frequencies contribute to $\gamma _{t}.$ On the other hand, from
Eq.$~$(\ref{PhasetimeApp}) we write%
\begin{equation}
\phi _{t}=\int_{0}^{\infty }d\omega J(\omega )\frac{\sin (\omega t)}{\omega
^{2}}.  \label{PhiThermal}
\end{equation}%
Using the spectral density$~$(\ref{OhmicSpectral}) it reads%
\begin{equation}
\phi _{t}=\frac{A}{2}(1-e^{-\gamma t}).
\end{equation}


\begin{thebibliography}{99}
\bibitem{EcoModulo} We note that $\mathrm{E}(\tau )$ is usually termed as
the LE amplitude while the absolute value $|\mathrm{E}(\tau )|^{2}$ is
termed as the LE. For simplicity, in the present contribution we denote $%
\mathrm{E}(\tau )$ as the LE.
\end{thebibliography}
\end{document}